\documentclass[twocolumn]{emulateapj}

\usepackage{graphics}
\usepackage{graphicx}
\usepackage{amsmath}

\graphicspath{{./}}

\shorttitle{Entrapment of CO in CO$_2$ ice}
\shortauthors{A. Simon et al.}

\begin{document}

\title{Entrapment of CO in CO$_2$ ice}

\author{Alexia Simon}
\affil{Harvard-Smithsonian Center for Astrophysics \\
60 Garden St, Cambridge, MA 02138, USA }

\author{Karin I. \"Oberg}
\affiliation{Harvard-Smithsonian Center for Astrophysics \\
60 Garden St, Cambridge, MA 02138, USA }

\author{Mahesh Rajappan}
\affil{Harvard-Smithsonian Center for Astrophysics \\
60 Garden St, Cambridge, MA 02138, USA }

\author{Pavlo Maksiutenko}
\affil{Harvard-Smithsonian Center for Astrophysics \\
60 Garden St, Cambridge, MA 02138, USA }

\begin{abstract}

Planet atmosphere and hydrosphere compositions are fundamentally set by accretion of volatiles, and therefore by the division of volatiles between gas and solids in planet-forming disks. For hyper-volatiles such as CO, this division is regulated by a combination of binding energies, and by the ability of other ice components to entrap. Water ice is known for its ability to trap CO and other volatile species. In this study we explore whether  another common interstellar and cometary ice component, CO$_2$, is able to trap CO as well.
We measure entrapment of CO molecules in CO$_2$ ice through temperature programmed desorption (TPD) experiments on CO$_2$:CO ice mixtures. We find that CO$_2$ ice traps CO with a typical efficiency of 40--60\% of the initially deposited CO molecules for a range of ice thicknesses between 7 and 50~ML, and ice mixture ratios between 1:1 and 9:1. The entrapment efficiency increases with ice thickness and CO dilution.
We also run analogous H$_2$O:CO experiments and find that
  under comparable experimental conditions CO$_2$ ice entraps CO more efficiently than H$_2$O ice up to the onset of CO$_2$ desorption at $\sim$70~K. We speculate that this may be due to different ice restructuring dynamics in H$_2$O and CO$_2$ ices around the CO desorption temperature. 
 Importantly, the ability of CO$_2$ to entrap CO may change the expected division between gas and solids for CO and other hyper-volatiles exterior to the CO$_2$ snowline during planet formation.

\end{abstract}


\section{Introduction} \label{sec:intro}

\ 

Planets and planetesimals acquire their volatiles through ice and gas accretion in protoplanetary disks. The division of volatiles between icy grain mantles and gas across the disk is a key ingredient for predicting atmosphere and hydrosphere compositions of planets forming at different disk locations \citep{Lewis74,Stevenson88,Oberg11e}. To a first approximation volatiles such as water, CO$_2$ and CO are distributed between ice and gas in disks based on their respective sublimation energies, and the temperature profiles of disks. Disk temperature profiles are mainly set by stellar radiation and by accretion heating \citep{dAlessio99}, resulting in a decreasing temperature with distance from the star. This temperature gradient should result in a series of condensation fronts or snowlines. Where a planet forms with respect to these snowlines will determine its initial composition, including the volatile elemental ratio, e.g. C/O/N ratios \citep{Piso16,Cridland16}.

In reality the division of volatiles between solids and gas is more complex than encoded in the snowline concept. Hyper-volatiles such as CO, N$_2$ and Argon can become entrapped in less volatile ice, especially in water ice, and therefore reside in disk solids much closer to the star than their sublimation energies would suggest \citep[e.g.][]{Bar-Nun85,Visser09,Marty17}. Such entrapment has been used to explain volatile abundances across the Solar System, including
C and N enrichment in Jupiter \citep{Owen99}, and CO and N$_2$ abundances in comets \citep{Notesco96,Notesco97,Bar-Nun07,Mousis12,Lectez15,Rubin15}. Experiments on entrapment kinetics together with cometary hypervolatile abundances have also been used to constrain whether cometary ices are interstellar or formed in the Solar Nebula \citep{Notesco03,Mousis16}. Entrapment of volatile radicals in H$_2$O ice may also be key to grow chemical complexity in interstellar and circumstellar ices \citep{Garrod08,Fresneau14}. 

So far, 'entrapment' has been used to denote any mechanism through which hyper-volatiles are retained in an ice matrix beyond their normal sublimation temperatures. There are at least two distinct entrapment mechanisms, however, `mechanical' entrapment and clathration \citep[e.g.][]{Hersant04}, which are active under different interstellar and nebular conditions. In short, 'mechanical entrapment' (from now on, simply entrapment) occurs when hyper-volatiles are trapped in pores in amorphous ice. By contrast, clathrate formation requires crystalline water ice, in which cages are stabilized by the inclusion of the hyper-volatile in lattice structure \citep[e.g.][]{Lunine85}. While the latter mechanism requires water ice, the former could in theory work for any amorphous ice with a porous structure. 

Even though entrapment could take place in any amorphous ice, the focus has been on water ice, the most abundant ice constituent in the interstellar medium, around low and high-mass protostars \citep{Gibb04,Oberg11c,Boogert15}, and in comets \citep{Mumma11}. Based on numerous experiments, amorphous water ice can indeed trap volatiles \citep[e.g.]{Bar-Nun85,Hudson91,Collings03b,Loeffler06,Fayolle11a}, and this may explain why e.g. some comets are rich in CO despite multiple past heating events above the CO sublimation temperature \citep{Mumma11}. 

Water is not the only abundant ice constituent, however. CO$_2$ is typically present at a 20-40\% abundance with respect to water in interstellar, circumstellar and cometary ices alike \citep{Boogert15,Mumma11}. In interstellar and circumstellar ices, a majority of the CO$_2$ is mixed in with the water ice, but some is also present in a separate, CO and CO$_2$ dominated ice phase \citep{Pontoppidan08}. The ice morphology in comets is largely unknown since ices are not observed directly, but for two comets 81P/Wild 2 and 67P/Churyumov-Gerasimenko, there is evidence for (partially) separate water and CO$_2$ ice phases. In the case of 81P/Wild 2, CO$_2$ and water outgassing appeared from different cometary locations, indicative of that some portion of the CO$_2$ is in a separate reservoir \citep{Mumma11}. During observations of 67P/Churyumov-Gerasimenko, CO$_2$ outgassing was also not well correlated with water outgassing, and indeed often anti-correlated \citep{Hoang17,Gasc17}. While different sublimation temperatures of H$_2$O and CO$_2$ may contribute to this behavior it unlikely explains it completely, since in laboratory experiments substantial amounts of CO$_2$ co-desorb with the water ice if the the two are originally mixed together \citep[e.g.][]{Fayolle11a}. Furthermore, the CO$_2$ outgassing did correlate with the outgassing of CO  \citep{Hassig15}, which raises the possibility that the observed CO was trapped in the CO$_2$ ice.

Based on  these clues we have undertaken an experimental study into the entrapment of CO in CO$_2$ ice, starting with amorphous CO$_2$:CO ice mixtures. \S2 presents the experimental apparatus and methods. In \S3 we first describe the outcome of 20 CO$_2$:CO ice experiments, and how entrapment depends on the initial ice thickness and ice mixing ratio. We then present results from a smaller set of analogous H$_2$O:CO experiments and compare the entrapment efficiencies in H$_2$O and CO$_2$ ice. \S4 discusses the experimental trends, possible origins for differences between CO$_2$ and H$_2$O entrapment efficiencies, and implications for volatile compositions across planet forming disks. \S5 presents some concluding remarks.

\section{Experimental methods}

\subsection{Experimental set-up}

The experimental set-up consists of a 6.5\arcsec ~ID ultra-high vacuum (UHV) chamber with a base pressure of 4$\times 10^{-9}$ Torr at room temperature. The chamber is designed to investigate thermal processes in interstellar ice analogs. The chamber is evacuated using a turbomolecular pump (Pfeiffer Vacuum Inc. HiPace® 300 with pump speed of 260 l/s for N$_2$) backed by a rotary vane pump (Pfeiffer Adixen 2005 SD Pascal) with an oil mist filter. The chamber pressure is monitored by MKS series 999 Quattro multi-sensor vacuum transducer coupled with a series PDR900 single channel controller which provides a continuous measurement from 10$^{-10}$ Torr to atmosphere. The UHV chamber is placed inside the sample compartment of a standard Fourier Transform Infrared (FTIR) spectrometer (Bruker Optics Inc. Vertex 70) with a liquid N$_2$ cooled Mercury Cadmium Telluride (MCT) detector. The IR spectrometer and the UHV chamber are placed on separate solid frames equipped with vibration isolation slides such that the optical path of the IR instrument is aligned with the horizontal axis of the chamber in the transmission geometry. The UHV chamber is separated from the spectrometer through two differentially pumped IR transparent KBr windows on either side. The entire optical path of the IR chamber is continuously purged with dry CO/CO$_2$ free air using a Parker 75-62 purge gas generator. 

The UHV chamber houses an IR transparent 2mm thick CsI window (0.75\arcsec ~clear view) mounted on an optical ring sample holder attached to a closed cycle helium cryostat (Advanced Research Systems (ARS) model DE-204S). The cryostat is integrated with DMX-20B interface which makes use of helium exchange gas to decouple the CsI window from the cold tip of the cryocooler, which prevents almost all vibrations from being transferred to the CsI window. The cryostat expander is also isolated from the sample mount via isolation bellows, which work in conjunction with gas heat exchange interface to avoid direct transmission of vibration to the IR spectrometer as well. The nickel plated copper radiation shield attached to the first stage of the cryostat helps to achieve CsI substrate temperature down to 17K. Note that because the shield is only cooled using the first stage it is too warm for very volatile ices such as CO to freeze out on it, while less volatile ices such as H$_2$O may condense there if an ice deposited through background deposition (not used in the experiments presented here).  The substrate temperature is controlled between 17K and 350K with a relative precision of +/-0.1K by a 75W resistive heater and a calibrated Si diode sensor (LS-670B-SD) located on the cryocooler tip near the heater using an external temperature controller (Lakeshore Model 335). A second identical Si diode sensor is mounted directly on the CsI substrate to measure the actual ice temperature. The absolute temperature uncertainty is estimated to 2--3~K. 

The differentially pumped rotary seal (Thermionics Vacuum Products, RNN-400) on the cryostat assembly allows 360 degree rotation of the CsI window inside the chamber during experiment. Ices are grown on the crycooled CsI window through directed vapor deposition using a gas dozer comprising a 4.8 mm ID stainless steel tube attached to a VAT variable leak valve and compact z-stage (MDC Vacuum Products, LLC). A stainless steel gasline is used to store gases and prepare gas mixtures of interest and deliver them into the chamber using the gas doser. The gasline is pumped down to pressures $<$5 x 10$^{-4}$ Torr using a turbopump station (Pfeiffer Vacuum Inc. HiCube Eco with pump speed of 67 l/s for N$_2$). The pressure in the gasline is monitored using two active capacitance transmitters (Pfeiffer Vacuum Inc.) CMR 361 and CMR 365 from $5\times10^{-4}$ to 1100 Torr. 

\subsection{Experimental procedures \label{sec:procedure}}

The experiments presented in this study are all temperature programmed desorption (TPD) experiments, in which a CO$_2$:CO or H$_2$O:CO ice mixture is heated using a linear heating ramp until desorption is complete. CO entrapment is measured by monitoring the desorption rate of CO during the ice warm-up using the QMS, and by monitoring the ice composition using the FTIR.

Gas mixtures are prepared inside the gasline attached to the chamber within an hour of the experiment. We also ran a few experiments where we let the gas-mixture sit overnight and found no significant difference in the experimental outcome. In the experiments, we used CO$_2$ (99\%, Sigma-Aldrich), CO (99\%, Sigma-Aldrich) and deionized H$_2$O liquid purified through multiple freeze-pump-thaw cycles.

The gas mixture is deposited onto the cryocooled substrate at 17~K until the desired ice thickness has been achieved. The latter is measured using infrared spectroscopy. In all experiments FTIR spectra are acquired at a spectral resolution of 1 cm$^{-1}$ from 4000 to 400 cm$^{-1}$. Each acquired spectra consists of 128 averaged scans and takes $\sim$3~min to complete. Background spectra are recorded before each ice deposition and subtracted from the respective ice spectra.

IR spectral band integrated opacities can be converted into ice column densities using the band strengths in Table \ref{tab:ices} from \citet{Hudgins93} and \citet{Gerakines95}, corrected for ice density in \citet{Bouilloud15}. We operate in transmission mode with the substrate at a 45$^\circ$ angle to the beam path. the calculation of ice column density is therefore

\begin{equation}
N_i =\frac{cos(\theta) \int \tau_i(v)dv}{A_i}
\end{equation}

where N$_i$ is the column density (molecule.cm$^{-2}$), cos($\theta$) is the angle between the IR field vector and the ice surface normal taking into account refraction at the vacuum-ice interface, $\int \tau_i(v)dv$ is the integrated optical depth of the IR band area, and $A_i$ is the band strength of the species. The column density in molecules cm$^{-2}$ is converted to monolayers of ice (ML) by using a standard 10$^{15}$ molecules cm$^{-2}$ conversion factor. We use the Eq. 1 both to calculate the initial ice matrix thickness (CO$_2$ or water) and the matrix:CO mixing ratio. The formal uncertainty in these calculations, based on the spectral rms and line widths, are $\sim$0.1~ML, but due to uncertainties in baseline subtractions we do not attempt to quantify ice thicknesses below 0.5~ML. The band strengths have reported uncertainties of 5--20\%, but this is for a specific temperature and ice mixture environment. In the case of CO, changes on the order of 10--30\% have been reported both when considering CO in different ice matrices at a single temperature, and for the same ice observed at different temperatures \citep[e.g.][]{dHendecourt86,Schmitt89,Gerakines95}. The reported ice mixture ratios and thicknesses are therefore only assumed to be accurate within 30\%. 

\begin{table}[!h]
\begin{center}
\caption{Integration ranges and band strengths (A) of CO, CO$_2$ and H$_2$O ices.}
\begin{tabular}{ cccc  }
  \hline\hline
Species & IR Band &  Band pos. (cm$^{-1}$) & A (cm) \\
 \hline 
CO & C-O str. & 2120-2170 & 1.4 x 10$^{-17}$ \\ 
CO$_2$ & C-O str. & 2320-2380 & 1.3 x 10$^{-16}$ \\ 
 H$_2$O & O-H str. & 3000-3600 & 2.2 x 10$^{-16}$ \\ \hline
\end{tabular}
\end{center}
\label{tab:ices}
\end{table}

\begin{figure*}[htp]
  \centering
  \includegraphics[width=0.3\textwidth]{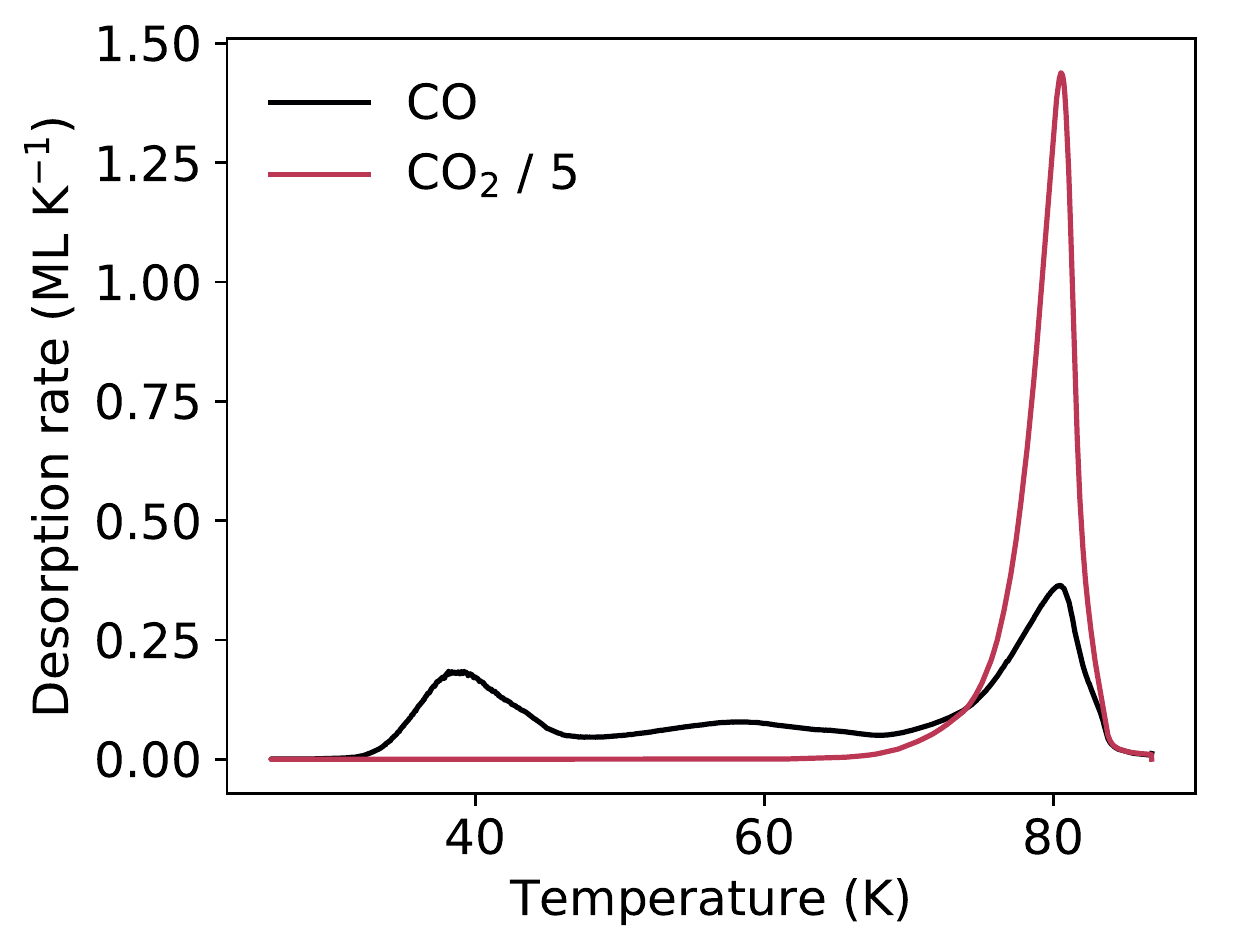}
  \includegraphics[width=0.35\textwidth]{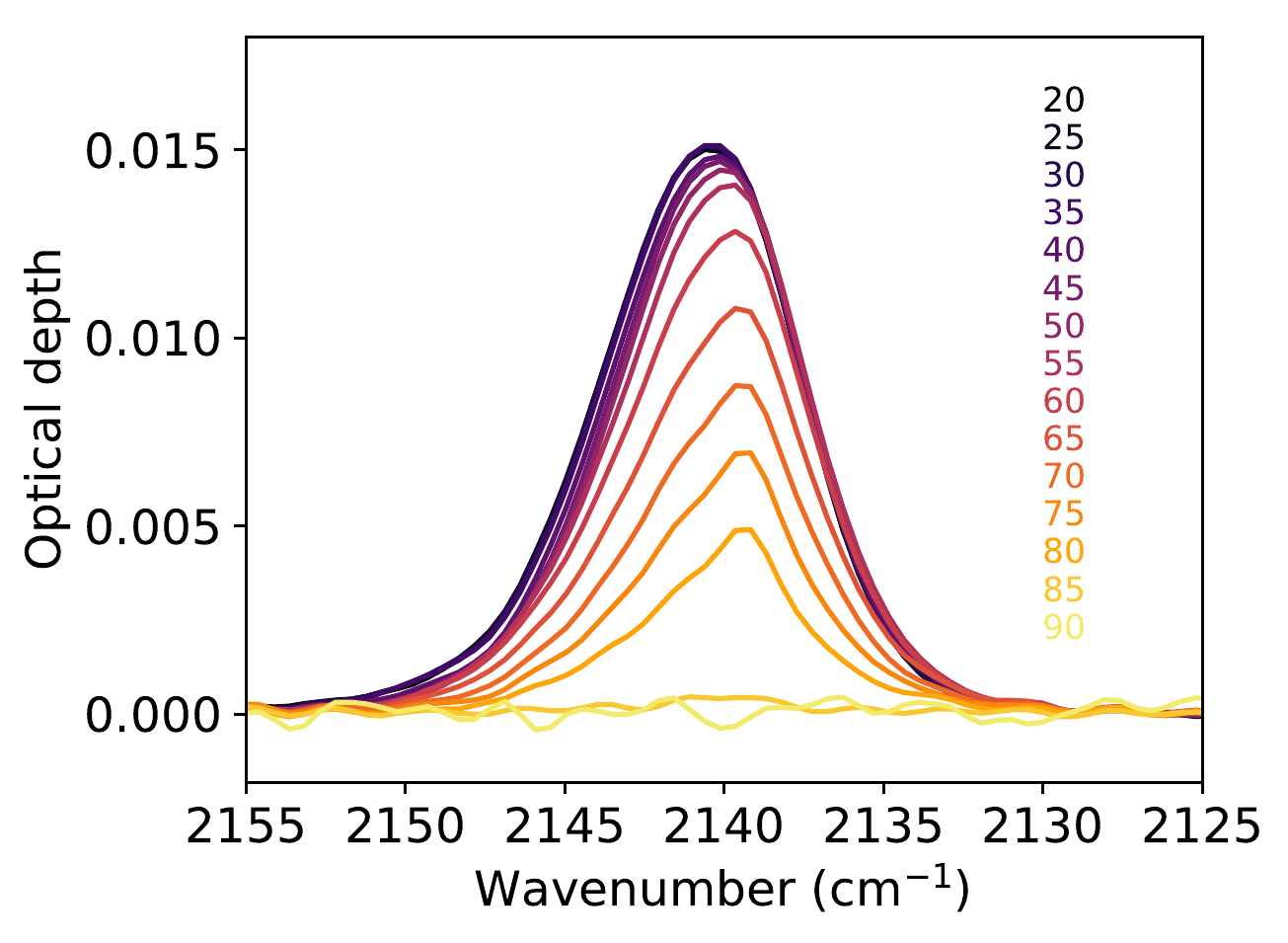}
  \includegraphics[width=0.3\textwidth]{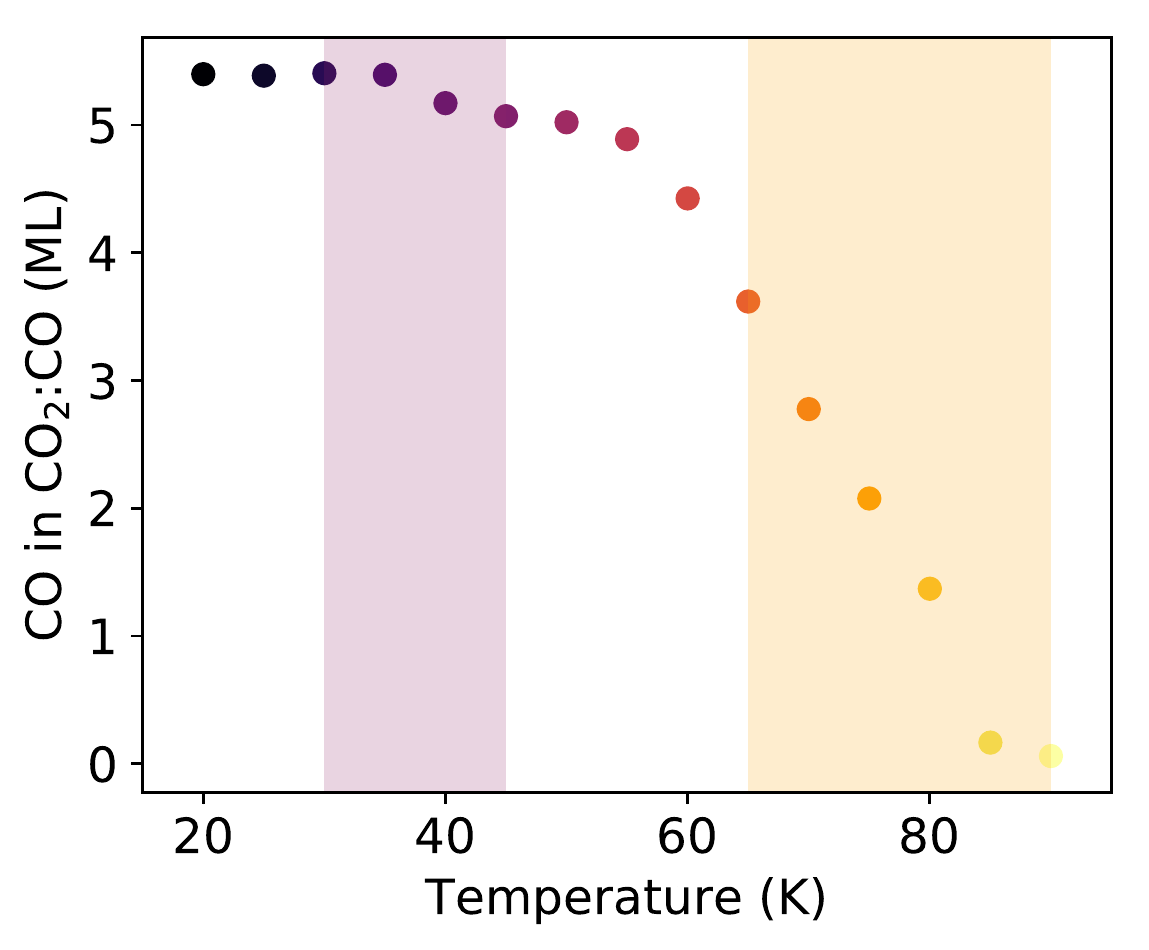}
 \caption{Experimental data used to measure CO entrapment. {\it Left:} TPD experiment of a CO$_2$:CO ice mixture (Exp. 16) showing CO desorption at the CO and CO$_2$ desorption temperatures, as well as an additional desorption feature centered at 55~K. {\it Middle:}  IR spectra of the CO band (2140 cm$^{-1}$) during the TPD. {\it Right:} Integrated IR CO band as a function of temperature. The CO and CO$_2$ desorption regions are colored purple and yellow, respectively. Less than expected CO loss is seen around the CO desorption temperature, probably due to ice re-structuring affecting the CO band strength. Formal uncertainties are $\sim$0.1~ML, similar to the symbol size.}
 \label{fig:example}
\end{figure*} 

Following ice deposition, the ice mixtures are heated with a constant rate of 1~K/min until complete desorption of the ice mixture in a so called temperature programmed desorption experiment (TPD). During the TPD, the ice desorption rate is measured every 1-2 seconds using the QMS, and IR spectra of the ice are acquired every 5~min. The shape and locations of TPD QMS curves encode information about the ice binding environment and structure, but in these experiments we only use the TPD curves to measure entrapment fractions based on the TPD characteristic that the integrated areas of a TPD curve between two time stamps is proportional to the number of molecules desorbed during that time.

The TPD curves are recorded as ion counts by the QMS, and this is converted into desorbing ML per K, by calibrating the total integrated TPD curve to the initial ice quantity measured by the FTIR. The smallest entrapment CO quantities we report with the QMS are 0.1~ML, since for smaller quantities the TPD baseline subtraction is uncertain. One complication that needs to be accounted for is that CO$_2$ partially fractionates into CO in the QMS. According to the NIST database, the strength of the signal is $\sim$10\% compared to the main m/z=44 signal. The exact electron impact energy of those measurements is not listed, however, and we also ran local tests, which were consistent with the NIST value. In all CO$_2$:CO TPD experiments 10\% of the CO$_2$ signal is therefore subtracted from the CO signal before any analysis.

\section{Results} \label{sec:results}

\subsection{CO entrapment measurements}

During TPD experiments we expect, based on previous experiments on H$_2$O:CO ice mixtures \citep[e.g.][]{Collings03b,Fayolle11a}, CO to desorb at two or more distinct temperatures corresponding to CO molecules escaping around the 'normal' CO desorption temperature of $\sim$30~K, and CO molecules that are trapped in the ice matrix. The latter CO may either be retained in the ice until the ice matrix desorbs around 80 and 150~K, for CO$_2$ and H$_2$O ice mixtures respectively, or escape during ice matrix restructuring during phase changes \citep[e.g.][]{Bar-Nun85,Hudson91,Collings03b,Collings04}. Figure \ref{fig:example} (left panel) shows that in the case of CO:CO$_2$ ice mixtures, most of the CO does indeed desorb either around 30~K, or together with CO$_2$ desorption.  
 In addition we observe a low level of CO desorption around 55~K, which we attribute to out-gassing of CO from the CO$_2$ ice matrix during CO$_2$ ice restructuring; a change in CO$_2$ ice spectral features has been observed around 60~K, attributed to the completion of CO$_2$ ice crystallization \citep{He2018}. 
 
  The fraction of trapped CO is obtained by integrating the CO QMS signal for temperatures $>$65~K, the earliest onset of CO$_2$ desorption in the experiments, and dividing it by the QMS signal integrated over the entire temperature range. In the experiment shown in Figure \ref{fig:example} the measured CO entrapment fraction is 51\%. 
  
 The QMS measurements are complemented by FTIR spectral time series that probe the CO ice quantity (Fig. \ref{fig:example}, middle and right panels).  The infrared spectra are converted to CO column densities in units of ML as described in \S\ref{sec:procedure}. Converting CO spectra into a CO quantity is complicated by a possible dependence of the CO ice band strength on ice morphology and composition, both which are expected to change during the warm-up ramp due to CO and ice matrix structural changes \citep[e.g.][]{Schmitt89,Isokoski13,He2018}.  
Figure \ref{fig:example} shows that indeed, while the QMS and FTIR CO measurements are broadly in agreement, there are quantitative differences. First, while there is a decrease in the CO ice IR feature strength at $\sim$30~K, corresponding to where one of the two major CO desorption peaks observed in Fig. \ref{fig:example} (left panel), the decrease in the IR is comparatively small. Second there is a sharper than expected decrease in the CO ice feature around 60~K. Third, for some CO$_2$:CO and all H$_2$O:CO entrapment experiments, the entrapment measured by the FTIR is considerably higher than those measured with the QMS; we estimate the FTIR CO entrapment fraction by dividing the CO ice abundance at 65~K (100~K in water ices) with the initial CO ice abundance. 

There are at least three possible explanations for this discrepancy: ice structure and temperature-dependent CO band strengths, diffusive ice build-up on the backside of the substrate, and uneven ice deposition  across the frontside of the substrate. The first two would affect the IR data, while the latter may result in an underestimate of entrapment using the QMS data. 

Band strengths are sensitive to ice structure and temperature. \citet{Schmitt89} found a substantial decrease in the CO band strength when heating up a previously annealed H$_2$O:CO mixture. This may account for the larger than expected decrease in CO intensity between 50 and 70~K in Fig. \ref{fig:example}. We also suspect that the band strength may increase around 30~K due to ice structure changes, which would explain the surprisingly small decrease in CO intensity during the first CO desorption peak, but this has not yet been tested experimentally.

While diffusive deposition of ice on the substrate backside could theoretically contribute to the IR/QMS discrepancy, we think it can be ruled out based on standard gas kinetic theory and the pumping rate and geometry of our system. We estimate that the ratio between backside and frontside deposition should be at most 5\%, i.e. very small. In set-ups where this is an important factor, IR measurements would underpredict entrapment since some of the measurements would be based on the thinner backside ice, where entrapment is likely less efficient.

Third, ice deposition on the frontside of the sample is expected to present a thickness gradient from the substrate center, since the gas is deposited through a narrow tube close to the substrate. This entails that while the IR probes an ice of almost constant thickness, the QMS will also measure entrapment fractions in the thinner ices deposited further from the substrate center on the frontside of the substrate, and on the sample holder. If this effect is important, it will result in a systematic underprediction of entrapment when using the QMS data.

With these uncertainties in mind, we record entrapment fractions using both methods. When exploring how entrapment fractions depend on ice thickness and compositions we mainly rely on the QMS data, which has a smaller experimental scatter. Since the systematic errors may be different for different ice matrix species we use both IR and QMS data when comparing entrapment in water and CO$_2$ ices.

\begin{table*}
\begin{center}
\caption{List of Experiments for CO$_2$:CO ices}
\label{tab:co2-co}
\begin{tabular}{ lcccccc  }
 \hline \hline
 No. &CO$_2$ &CO & Initial CO$_2$:CO  &CO${\rm _{Trapped}^*}$ & CO${\rm _{Trapped}^*} /$ &  Trapped CO$_2$:CO$^{**}$\\
 &(ML) &(ML)& mixture at T=20~K & (ML) &CO$\rm _{Initial}$ (\%)&mixture  at T=70~K\\
 \hline 
 1 & 6.7 & 6.6 & 1:1 & 0.94 / $<$0.5 & 14\% / $<$8\% & 7:1 \\ \hline
2 & 7.1 & 4.6 & 2:1 & 1.1 / 0.93 & 24\% / 20\% & 7:1\\
3 & 21 & 13 & 2:1 & 4.4 / 2.6  &  32\% / 20\% & 5:1\\
4 & 23 & 15 & 2:1 & 3.2 / 3.6  &  21\% / 24\% & 7:1\\
5 & 30 & 13 & 2:1 & 5.0 / 3.8 &  40\% / 31\% & 6:1\\
6 & 31 & 13 & 2:1 & 5.1 / 3.9 &  40\% / 30\% & 6:1\\  
\hline
7 & 32 & 10 & 3:1 & 4.6 / 4.5 & 45\% / 43\% & 7:1\\
8 & 53 & 20 & 3:1 & 9.3 / 8.2 & 46\% / 41\% & 6:1\\ 
\hline
9 & 5.3 & 1.3 & 4:1 & 0.57 / 0.69  &  44\% / 53\% & 11:1\\
10 & 8.2 & 2.3 & 4:1 & 1.1 / 1.6  & 46\% / 66\% & 9:1\\
11 & 24 & 6.2 & 4:1 & 3.6 / 4.0  &  55\% / 65\% & 7:1\\
12 & 31 & 7.1 & 4:1 & 4.0 / 4.0  &  57\% / 56\% & 8:1\\ 
13 & 37 & 9.5 & 4:1 & 4.8 / 4.9 &  50\% / 51\% & 8:1\\
14 & 42 & 11 & 4:1 & 6.5 / 6.2  &  60\% / 57\% & 6:1\\ 
\hline
15 & 16 & 3.1 & 5:1 & 1.5 / 2.1  &  49\% / 67\% & 11:1\\ 
16 & 29 & 5.5 & 5:1 & 2.8 / 3.7 &  51\% / 67\% & 10:1\\  
\hline
17 & 36 & 4.8 & 7:1 & 2.8 / 3. 8&  59\% / 80\% & 12:1\\ 
\hline
18 & 10 & 1.3 & 8:1 & 0.60 / 0.94  &  53\% / 74\% & 17:1\\ 
\hline
19 & 5.7 & 0.66 & 9:1 & 0.27 / $<$0.5  &  43\% / $<$71\% & 21:1\\
20 & 30 & 3.4 & 9:1 & 2.1 / 3.0 &  60\% / 87\% & 14:1\\
 \hline
\end{tabular}
\\$^*$ X / Y denotes CO entrapment according the QMS data (X) and FTIR data (Y). QMS uncertainties are $<$0.1~ML, and formal FTIR uncertainties are $\sim$0.1~ML, but this does not take into account probable band strength variations during the experiment.
\\$^{**}$ Based on QMS measurement.
\end{center}
\end{table*}

\subsection{CO$_2$:CO Experiments \label{sec:res-co2}}

Informed by observed interstellar ices, the CO$_2$:CO experiments include CO$_2$ ice thicknesses between 7 and 53~ML, and CO$_2$:CO mixing ratios between 1:1 and 9:1 \citep{Pontoppidan08}. To disentangle the effects of ice thickness and composition, most ice mixtures were run with a CO$_2$ ice thickness of $\sim$30--35~ML, and one ice mixture (4:1) includes an order of magnitude range in ice thicknesses. Table \ref{tab:co2-co} lists all CO$_2$:CO entrapment experiments, organized by CO$_2$ ice thickness, in ML, then by CO ice thickness for each ice mixing ratio.

\begin{figure}[h!]
  \centering
  \includegraphics[width=0.45\textwidth]{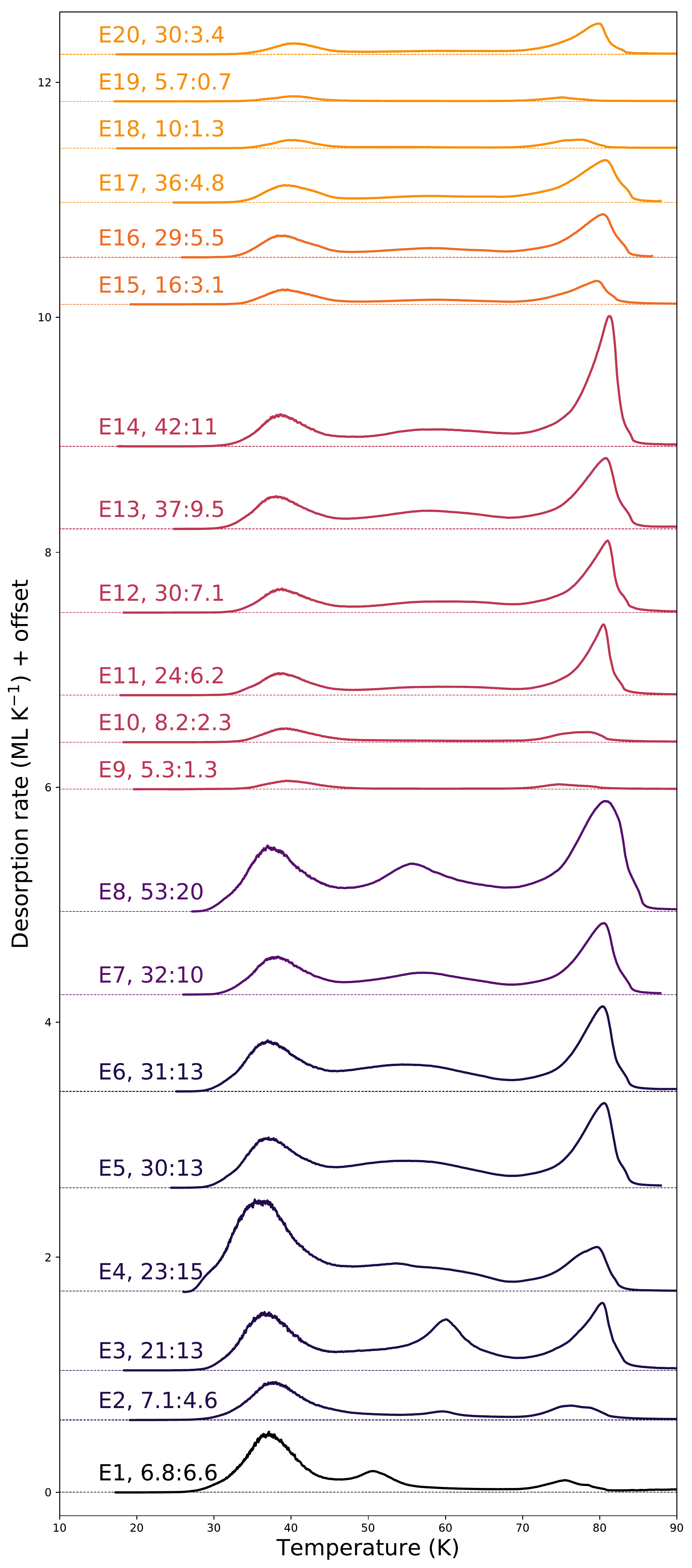}
 \caption{Temperature Programmed Desorption (TPD) data of CO desorption from the CO$_2$:CO ices listed in Table \ref{tab:co2-co}.}
 \label{fig:co2-tpd}
\end{figure}

Figure \ref{fig:co2-tpd} presents the CO TPD spectra  listed in Table \ref{tab:co2-co} in the same order. All TPD spectra show two major CO desorption peaks corresponding to CO desorbing from the ice surface, and co-desorption of entrapped CO at the CO$_2$ desorption temperature, and generally also some desorption at intermediate temperatures as discussed in the previous section. In most experiments the intermediate peak occurs above 50~K, but in the thinnest and only 1:1 mixing ratio experiment it occurs already at 50~K, possibly because of a different ice re-structuring dynamics in this highly concentrated ice. The relative size of the two major CO desorption peaks vary in the different experiments; in TPD experiments with similar ice mixing ratios, the ratio of the entrapment peak to the CO desorption peak increases with ice thickness, and for experiments with similar CO$_2$ ice thickness, the ratio increases with decreasing CO concentration. The relative size of the intermediate peak decreases with ice thickness and with dilution.

CO entrapment in CO$_2$ ices is quantified in Table \ref{tab:co2-co} using both the QMS and FTIR approaches. Using the QMS data, the CO trapping fraction varies between 14 and 60\% with respect to the initial CO content. For the majority of CO$_2$:CO experiments, the entrapment efficiency is $>$40\%, and for all but 3 experiments it is $>$30\%. The only experiments with a low ($<$30\%) entrapment efficiency are both highly concentrated in CO (CO$_2$:CO mixing ratios of $<$2), and thin ($<$25~ML of CO$_2$). 

Table \ref{tab:co2-co} can also be used to estimate the entrapment measurements error. One of the experiments is exactly replicated (Exp 5 and 6), and the resulting difference is 2\% in the quantity of trapped CO. If Exp. 7 is also considered a duplicate (it almost is), the error increases to 4\% for the QMS measurements and 7\% for the IR measurements. The QMS uncertainty is added in quadrature to 0.1~ML, the fundamental TPD uncertainty, to all QMS entrapment measures in the subsequent analysis. 

\begin{figure}[h!]
  \centering
  \plotone{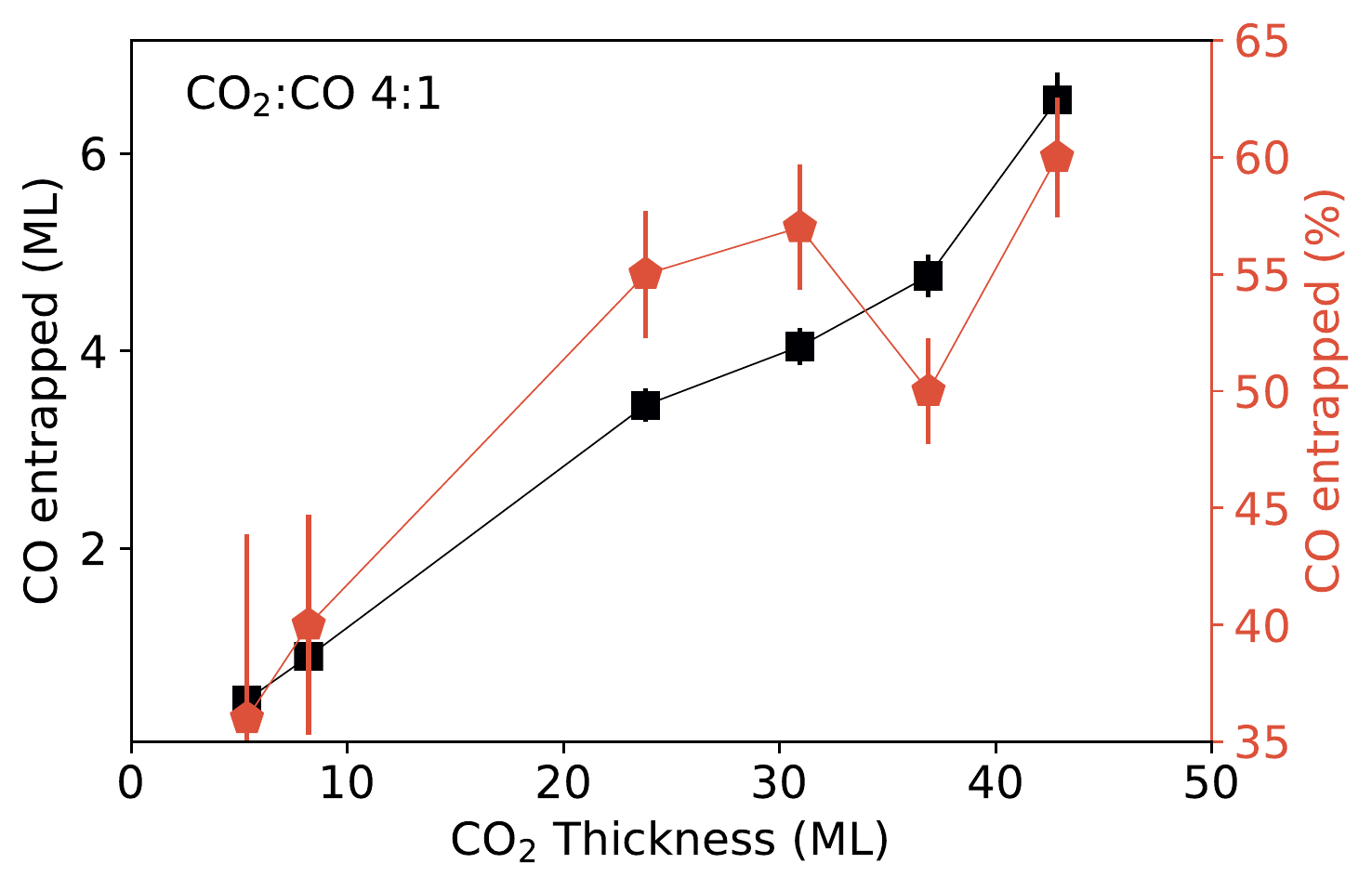}
  \plotone{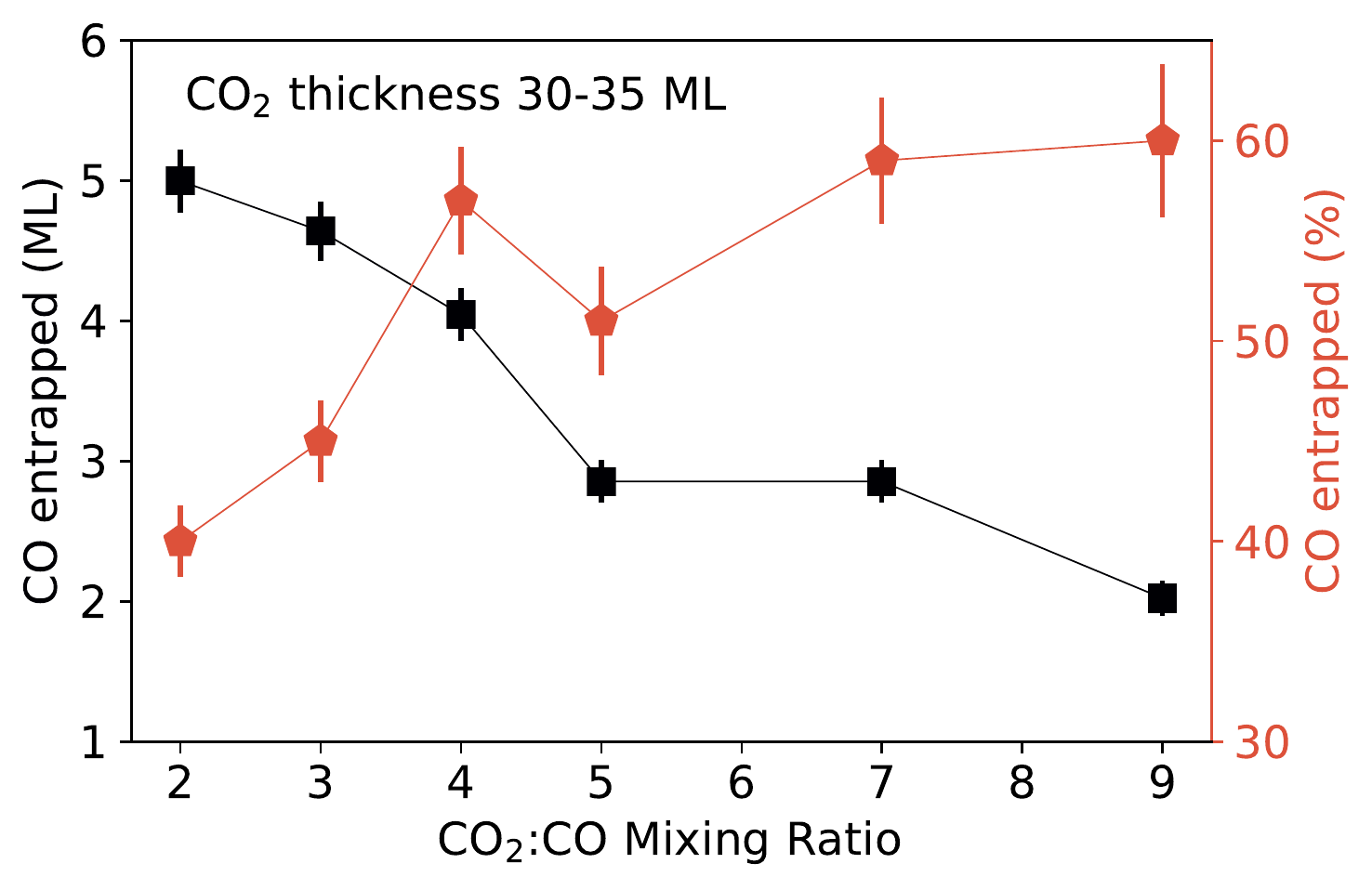}
 \caption{Experimental results for CO entrapment depending of the thickness and ratios. {\it Upper panel:} Entrapment efficiency for CO$_2$:CO, 4:1 mixing ratio at different thicknesses. {\it Lower panel:} Entrapment efficiency in CO$_2$:CO ices with CO$_2$ ice thicknesses of 30--35ML and different mixing ratios.  }
 \label{fig:co2-trap}
\end{figure}  

\begin{figure}[h!]
  \centering
  \plotone{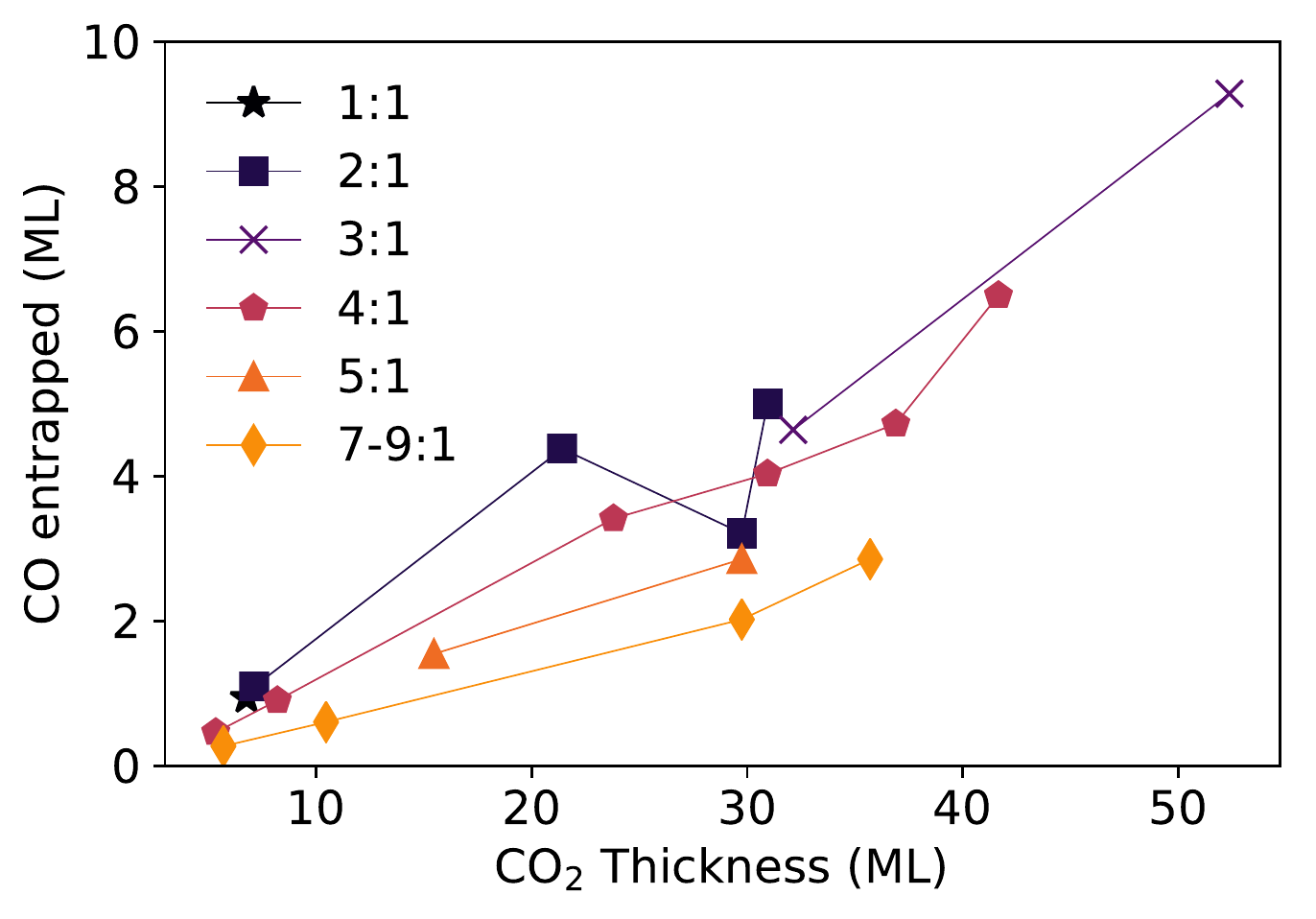}
  \plotone{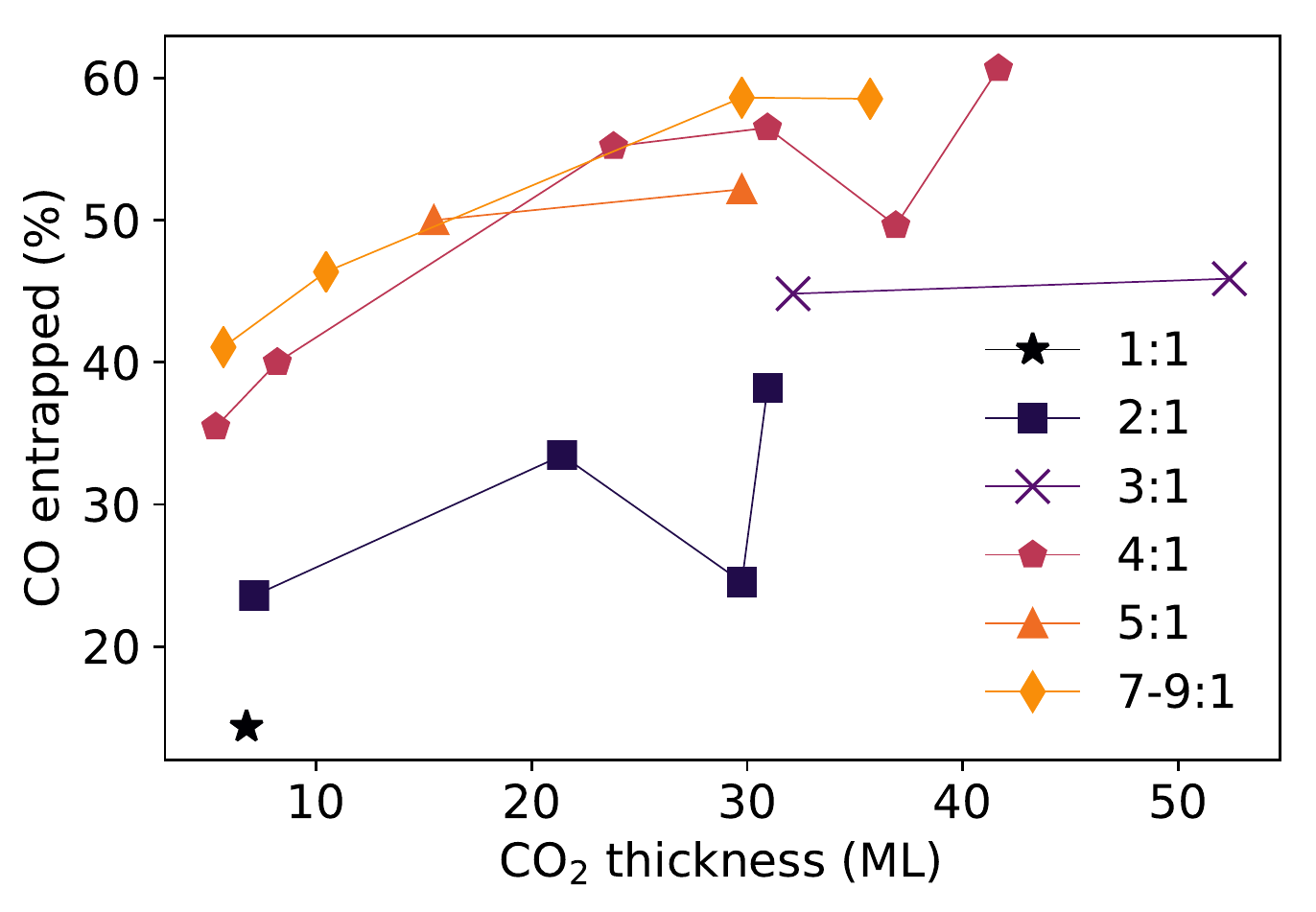}
 \caption{CO entrapment dependencies on CO$_2$ ice thickness and CO$_2$:CO ice mixing ratio for all experiments listed in Table \ref{tab:co2-co}.  {\it Upper panel:} Quantity of CO entrapped as a function of CO$_2$ ice thickness for ice mixing ratios between 1:1 and 7--9:1 (symbols and colors). {\it Lower panel:} Fraction of the initial CO that is trapped, otherwise as upper panel. Error bars are not shown for clarity, and are comparable to Fig. \ref{fig:co2-trap}. }
  \label{fig:co2-all}
\end{figure}  

Figure \ref{fig:co2-trap}, upper panel illustrates how the CO entrapment efficiency depends on ice thickness for a single ice mixing ratio (4:1). When considering the quantity of trapped CO, there is a monotonic increase with CO$_2$ ice thickness between 7 and 50~ML. However, the entrapment efficiency, in \% of the original CO, levels out above an initial CO$_2$ ice thickness of 20~ML.

 The mixing ratio also influences CO entrapment, which is exemplified in the lower panel of Figure \ref{fig:co2-trap} for CO$_2$:CO ices with initial CO$_2$ ice thickness of 30--35~ML. The quantity of CO trapped (black line) decreases with increasing mixing ratio. This is a trivial result, however, since if the ice thickness is constant and the mixing ratio increases there is simply less CO to start with in the ice. By contrast,  the fraction of CO trapped increases with the mixing ratio between 2:1 and 4:1, before leveling out. 

Figure \ref{fig:co2-all} shows the relationship between entrapment and CO$_2$ ice thickness for all different mixing ratios between 1:1 and 9:1 (Exp. 4 is excluded since it is right in between 1:1 and 2:1). For convenience the 7:1, 8:1 and 9:1 experiments are grouped together. The trends observed for 4:1 ice mixtures and different ice mixtures for CO$_2$ ice thicknesses of 30--35~ML appear to hold for the full set of experiments.  That is, the quantity of trapped CO increases monotonically with CO$_2$ ice thickness, and with increasing CO concentration for a given ice thickness. The percentage of trapped CO increases with ice thickness and with decreasing CO concentration when the ice thickness is $<$20~ML and the CO$_2$:CO ice mixing ratio is $<$4:1. For thicker and more dilute ice mixtures, the CO entrapment efficiency levels off.

\begin{figure}[h!]
  \centering
  \plotone{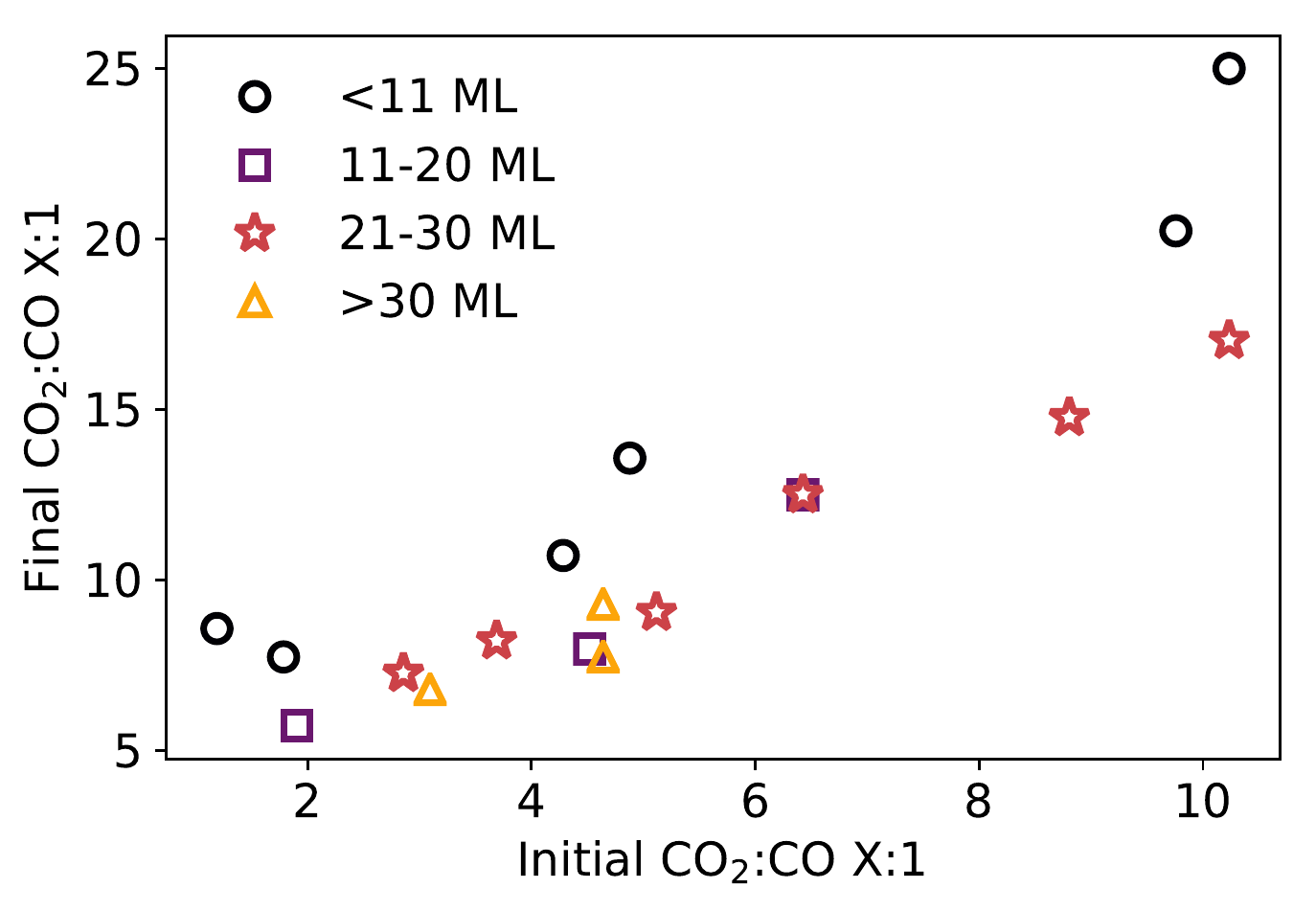}
 \caption{The relationship between initial and final or trapped CO$_2$:CO ice mixing ratios as a function of ice thickness. Typical mixing ratio uncertainties are $X^{+1}_{-1}$:1 and increase with CO dilution.}
  \label{fig:co2-final}
\end{figure}

In addition to entrapment efficiencies, we can also determine how the CO$_2$:CO ice mixing ratio for trapped CO, i.e. how many CO$_2$ molecules are present in the ice per trapped CO molecule at the onset of CO$_2$ desorption, depends on the initial ice thickness and mixing ratio. Inspecting Table \ref{tab:co2-co}, the entrapment mixing ratio appears to increase with increasing initial mixing ratio for similar ice thicknesses, while the impact of ice thickness is less clear. This is illustrated in Fig. \ref{fig:co2-final}, which shows a strong relationship between the initial and final mixing ratios, such that the most concentrated final CO$_2$ ices are achieved in the experiments with the initially most concentrated ices. The thinnest ices (CO$_2<10$~ML) show systematically higher CO$_2$:CO entrapment mixing ratios, indicative of less efficient entrapment in the thinnest ices, but there is no clear trend with thickness above 10~ML. The maximum trapped CO$_2$:CO mixing ratio observed in the experiments is 5:1, i.e. at least five CO$_2$ molecules is needed to trap one CO molecule in ices up to $\sim$30~ML in thickness.

\begin{table*}
\begin{center}
\caption{List of Experiments for H$_2$O:CO ices}
\begin{tabular}{ lcccccc  }
 \hline \hline
 No. & H$_2$O &CO & Initial H$_2$O:CO  &CO${\rm _{Trapped}^*}$ & CO${\rm _{Trapped}^*} /$ &  Final H$_2$O:CO$^{**}$\\
 &(ML) &(ML)& at mixture at T=20~K & (ML) &$CO\rm _{Initial}$& mixture at T=130~K\\
 \hline 
1 & 12 & 12 & 1:1 & 0.35 / 0.68 & 3\% / 6\% & 35:1 \\ \hline
2 & 6.8 & 4.1 & 2:1 & $<$0.1 / $<$0.5 & $<$2\% / $<$12\% & $>$58:1 \\
3 & 15 &	5.9 & 2:1 &	0.12 / $<$0.5 &	2\% / $<$8\% &	120:1 \\
4 & 20 &	9.9 & 2:1 &	0.50 / 1.6	& 5\%	/ 16\% &	41:1 \\
5 & 41 &	19 & 2:1 &		1.3 / 3.1	& 7\%	/ 17\% & 31:1 \\ 
6 & 61 &	35 & 2:1 &		3.2 / 6.0	& 9\%	/ 17\% & 19:1 \\ \hline
7 & 28 &	11 & 3:1 &		1.1 / 1.9	& 10\% / 	17\% & 26:1 \\
8 & 30 &	9.2 & 3:1 & 	1.1 / 2.1	& 12\% / 	23\%	& 27:1 \\ \hline
9 & 18 &	3.3 & 5:1 &	0.60 / 1.8	& 18\% /	55\%	& 30:1 \\ \hline
11 & 28 & 2.8 & 10:1 &	0.70 / 1.9	& 26\% /	66\%	& 40:1 \\
10 & 28 & 2.7 & 10:1 &	0.96 / 1.5	& 34\% / 	59\%	& 29:1 \\ 
 \hline
\end{tabular}
\label{tab:h2o-co}
\\$^*$ X / Y denotes CO entrapment according the QMS data (X) and FTIR data (Y).
\\$^{**}$ Based on QMS measurement.
\end{center}
\end{table*}

\subsection{H$_2$O:CO Experiments}

\begin{figure}[h!]
  \centering
  \plotone{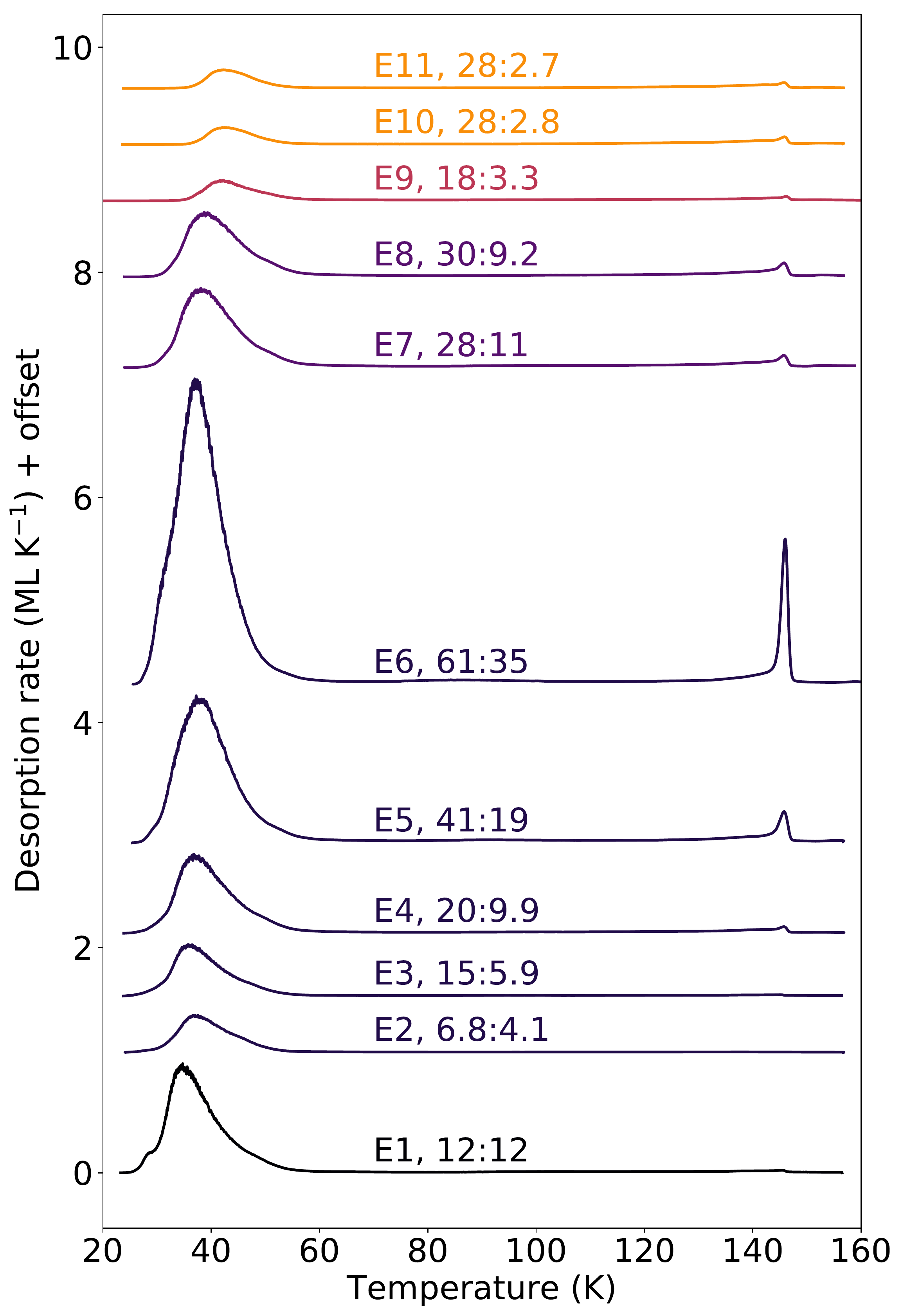}
 \caption{CO desorption from the H$_2$O:CO ices during TPD experiments listed in Table \ref{tab:h2o-co}.}
 \label{fig:h2o-tpd}
\end{figure} 

Table \ref{tab:h2o-co} lists experiments on CO entrapment in H$_2$O:CO ices, analogous to Table \ref{tab:co2-co} for CO$_2$:CO experiments. H$_2$O:CO TPD experiments were run for a similar range of ice thicknesses and mixing ratios as CO$_2$:CO. The number of H$_2$O:CO experiments is, however, smaller, since their main purpose is to serve as a comparison set to the CO$_2$:CO experiments.
Figure \ref{fig:h2o-tpd} shows CO desorption in the H$_2$O:CO TPD experiments in the same order as listed in Table \ref{tab:h2o-co}. Similar to the CO$_2$:CO experiments, there are two desorption peaks, around 30--40~K, and around 150~K. The second peak slightly precedes the H$_2$O desorption peak, and based on previous experiments it is due to CO outgassing during H$_2$O ice crystallization \cite[e.g.][]{Collings03b}. By contrast to the CO$_2$:CO experiments, the first peaks always dominates, indicative of that only a small fraction of the initial CO is trapped. 


\begin{figure}[ht]
  \centering
  \plotone{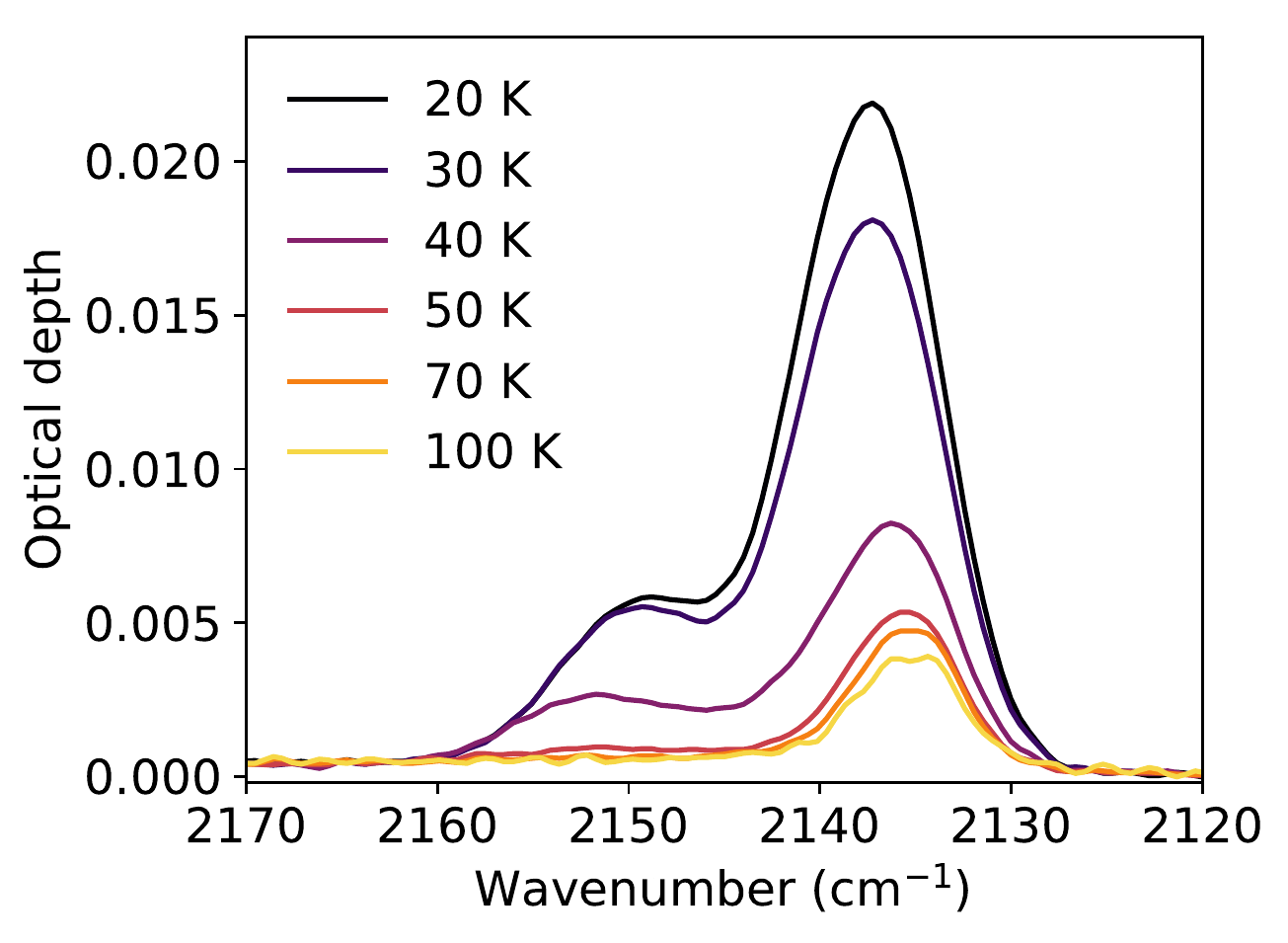}
 \caption{Infrared spectra of the CO IR feature during warm-up of a H$_2$O:CO 2:1 ice mixture, demonstrating the distinctly different CO ices found below and above 40~K.}
 \label{fig:h2o-IR}
\end{figure} 

\begin{figure}[ht]
  \centering
  \plotone{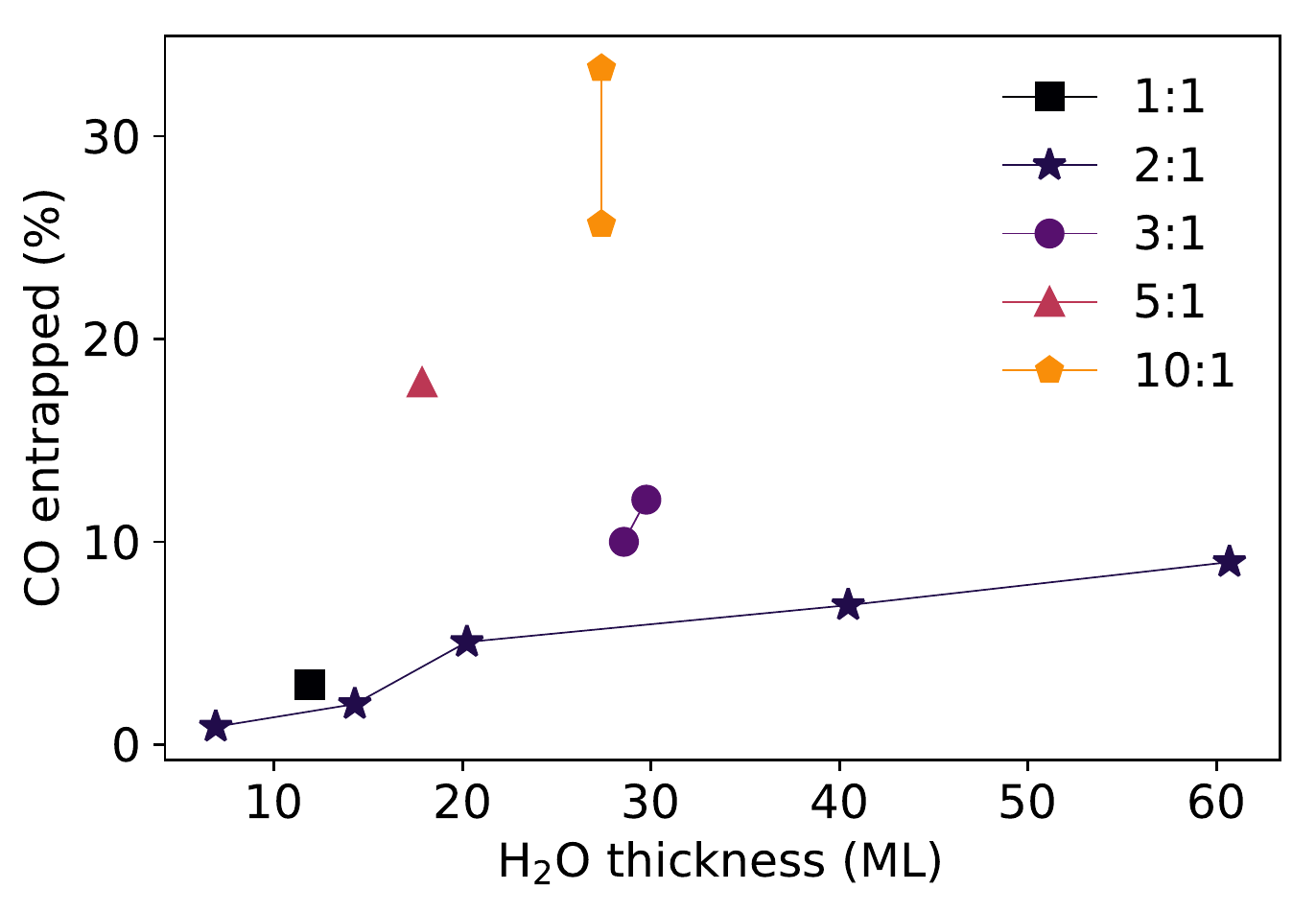}
 \caption{ CO entrapment dependencies on H$_2$O ice thickness and H$_2$O:CO ice mixing ratio for H$_2$O ice thickness between 7 and 60~ML, and ice mixing ratios between 1:1 and 10:1.}
 \label{fig:h2o-all}
\end{figure}

The CO entrapment fractions in the H$_2$O:CO experiments are listed in Table \ref{tab:h2o-co} and illustrated in Fig. \ref{fig:h2o-all}. Compared to the CO$_2$ experiments there is a larger discrepancy between QMS and FTIR results; the FTIR time series suggests on average a factor of 2 higher entrapment fraction. Similar to the CO$_2$:CO experiments, the spectra of CO in cold and warm H$_2$O ice are distinctly different (Fig. \ref{fig:h2o-IR}), suggesting that the CO band strengths may also be different in the two temperature regimes, but as discussed above, the QMS data could also be underestimating the entrapment fraction. According to the QMS measurements the CO entrapment efficiency is $<$10\% in a majority of experiments. While the entrapment is higher when basing the analysis of the IR measurements, the entrapment efficiency is still low, $<$25\%, in all but the three most dilute ices. This can be compared to the CO$_2$:CO experiments where only a small fraction of thin and highly concentrated ice mixtures resulted in a similarly low CO entrapment efficiency. 
The CO entrapment trends are similar to those observed in CO$_2$:CO ices, i.e. the entrapment efficiency increases with ice thickness and decreasing CO concentration, which is also in qualitative agreement with previous studies on CO:H$_2$O ices \citep{Hudson91,Fayolle11a}.

\subsection{CO entrapment in CO$_2$ vs. H$_2$O ices}

\begin{figure}[h!]
  \centering
  \includegraphics[width=0.4\textwidth]{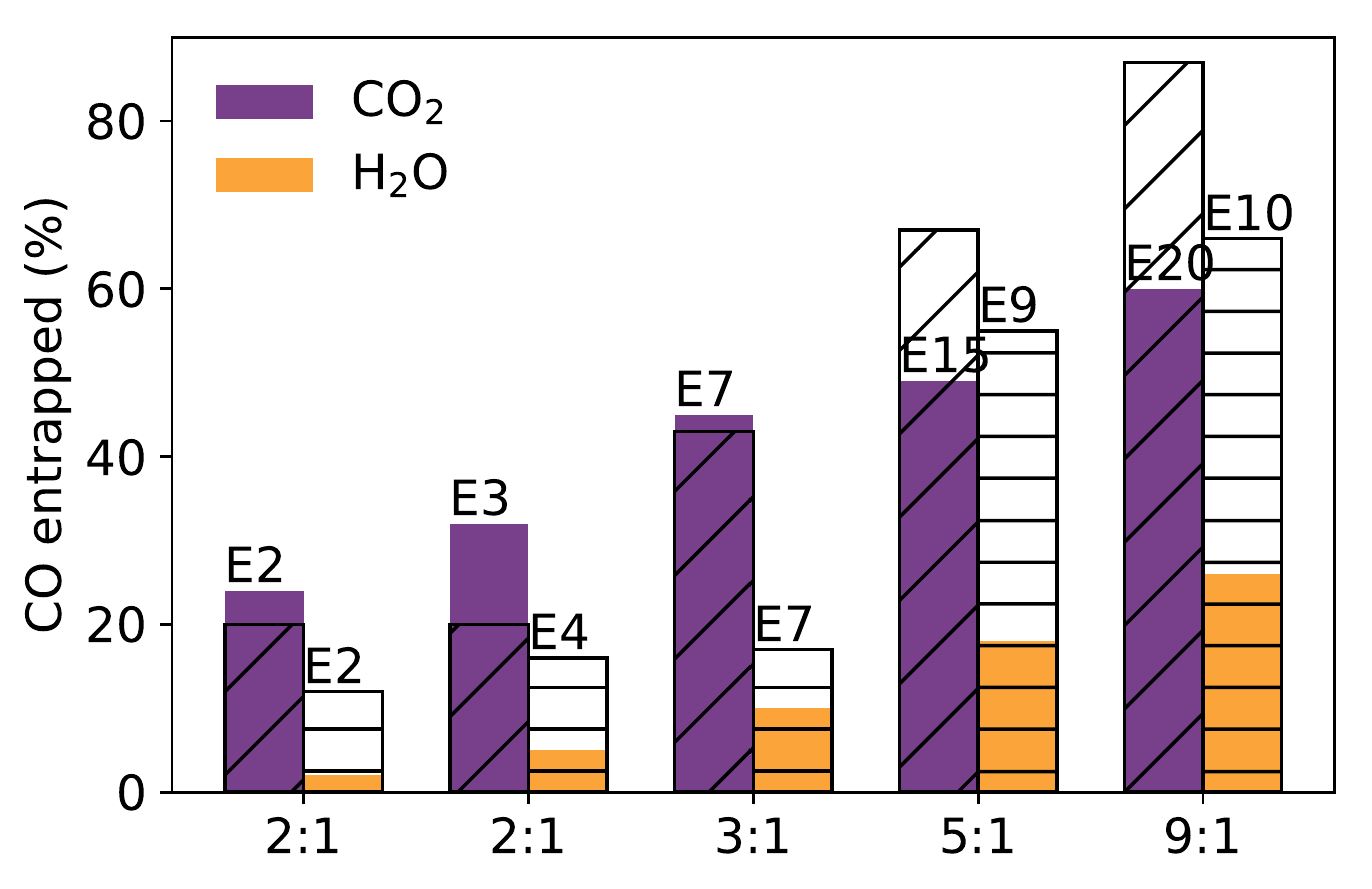}
 \caption{Comparison of CO$_2$ and H$_2$O ice entrapment efficiencies of CO as a function of ice mixture ratio. The ice matrix thicknesses in the five comparisons are $\sim$7, 21, 30, 16, and 29~ML, respectively. The solid bars use the QMS TPD data and the hatched bars the IR time series to estimate entrapment fractions.}
 \label{fig:comp}
\end{figure}

Figure \ref{fig:comp} shows a direct comparison of CO entrapment efficiencies in CO$_2$ and H$_2$O ices. Five of our CO$_2$:CO ice experiments are well matched to H$_2$O:CO experiments in terms of mixing ratio and ice matrix thickness. These are experiments number 2, 3, 7, 15 and 20 in Table 2, and 2, 4, 7, 9 and 10 in Table 3. These experiments are compared pairwise in Fig. \ref{fig:comp}, though note that CO is considered trapped if it is present above 65~K in CO$_2$ ice and above 100~K in H$_2$O ice. Solid bars compare entrapment using the QMS data, and in all cases, substantially more CO is retained in the CO$_2$ ices compared to the analogous H$_2$O ices. The difference is the largest for the thinnest and most concentrated ices, where CO$_2$ traps up to an order of magnitude more of the initial CO. For thicker and more dilute ices the difference is smaller, but CO$_2$ still appears to be $>2\times$ more efficient than H$_2$O in retaining CO. The decreasing gap between CO$_2$ and H$_2$O entrapment efficiencies for more dilute ices, indicates that the efficiency of H$_2$O to entrap hyper-volatiles is more sensitive to specific ice characteristics compared to CO$_2$. We also compare entrapment using IR time series (hatched bars), and while CO$_2$ is still the more efficient entrapper, the difference between water and CO$_2$ is smaller: CO$_2$ traps an extra 20-250\% CO compared to water in matched experiments.

\section{Discussion}

In the previous section, we showed that CO is efficiently trapped in CO$_2$ following warm-up of CO:CO$_2$ ice mixtures from 20~K to CO$_2$ desorption. Surprisingly, this entrapment efficiency is higher than in analogous H$_2$O:CO ice experiments. Among the CO$_2$:CO ice experiments the entrapment efficiency increases with ice thickness and dilution, but only up to a point. The origin and implications of these trends are discussed in \S\ref{disc-trends}. Possible explanations for the higher entrapment efficiency of CO$_2$ compared to H$_2$O are presented in \S\ref{disc-h2o}, and the astrophysical implications for the discovered CO$_2$ entrapment are introduced in \S\ref{disc-astro}.

\subsection{CO  entrapment in CO$_2$ ice \label{disc-trends}}

The observed CO$_2$ entrapment characteristics and trends reported in \S\ref{sec:res-co2} encode information on under which conditions CO$_2$ ice can retain mixed in CO molecules, and on the entrapment mechanism. First, the entrapment efficiency increases with ice thickness, but only for thin ices, $<$10--30~ML. For thicker ices, the entrapment efficiency is constant. This behavior implies that the entrapment efficiency decreases towards the ice surface in the upper ice layers. Deeper into the ice, CO$_2$ appears to retain a constant fraction of the initially mixed in CO. The  CO$_2$ ice thickness required to reach this constant efficiency increases with CO concentration, indicating that CO  out-gassing is possible from deeper into the ice in more concentrated ices.

Regardless of ice thickness, there seems to be a minimum required number CO$_2$ molecules per trapped CO molecule. Based on the observed entrapped CO$_2$:CO mixing ratios this number is approximate 5 CO$_2$ molecules per trapped CO molecule.
 This can be compared e.g. to e.g. studies of H$_2$O clathration, where a similar proportionality is required, suggesting that the two entrapment mechanisms can result in similar number of hyper-volatiles in same-sized ice matrices \citep[e.g.][]{Lunine85}.

\begin{figure}[h!]
  \centering
  \includegraphics[width=0.4\textwidth]{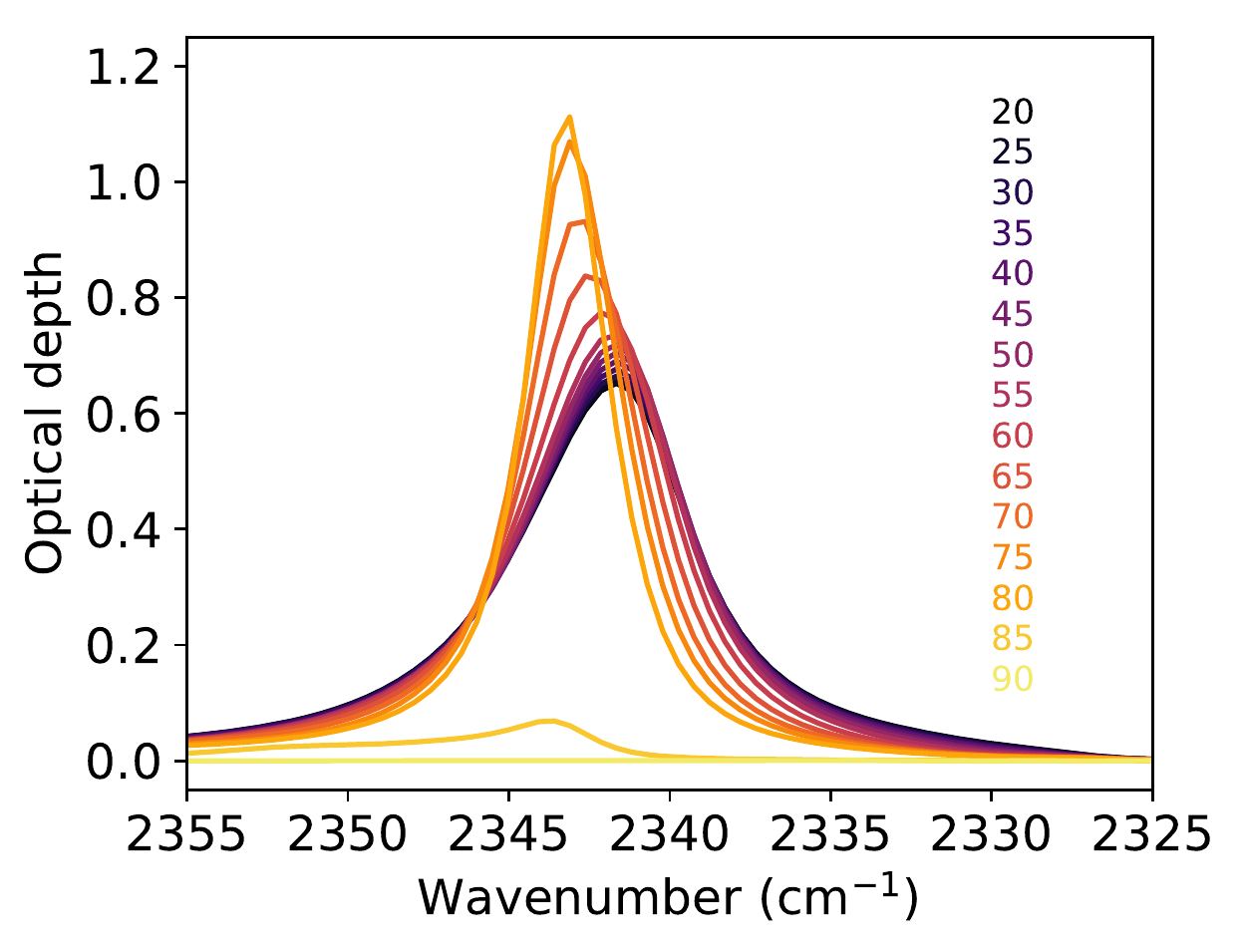}
 \caption{Spectra of the CO$_2$ ice band during warm-up in Exp. 16. The ice temperatures in K the different spectra were acquired at are shown in the legend.}
 \label{fig:co2-sp}
\end{figure}

A second set of constraints on the entrapment mechanics comes from the TPD spectra and FTIR time series. Both  show evidence for substantial CO loss in between the CO and CO$_2$ desorption temperatures. This region coincides with a thickness-dependent completion of CO$_2$ crystallization \citep{He2018}. This outgassing of CO is reminiscent of the previously observed outgassing of CO from H$_2$O during H$_2$O crystallization \citep{Bar-Nun85,Collings03b}. The results of this phase change is seen both in CO and CO$_2$ spectral shapes. In Fig. \ref{fig:example}, the CO band becomes more structured between 50 and 75~K, indicative of that the the ice environment within which CO resides is changing. Concurrently, there is a clear shift and narrowing of the CO$_2$ ice band (Fig. \ref{fig:co2-sp}).

Finally, spectroscopy of the CO$_2$ ice feature during ice warm-up (Fig. \ref{fig:co2-sp}) shows that CO$_2$ indeed has undergone some restructuring already by 30~K, similar to what was observed for pure CO$_2$ ice \citep{He2018}. If CO is trapped through pore collapse, analogous to what has been proposed for amorphous water ice \citep{Collings03b}, then the spectroscopy suggests that CO$_2$ mobility and therefore CO$_2$ pore collapse may begin prior to the onset of CO desorption, which may help to explain why it can trap CO molecules so efficiently.
It thus appears that CO$_2$ ice restructuring holds a key to both CO retention and CO escape from CO$_2$ ices.

The scenario emerging from the above constraints is the following: At $\sim$30~K, CO becomes volatile in the ice and begins to escape from the CO$_2$ ice surface and pores. Around the same temperature CO$_2$ becomes mobile and begins to crystallize, trapping CO with an increasing efficiency with ice depth. Between 40 and 60~K, CO$_2$ crystallization completes and during this process some of the remaining CO escapes as the CO$_2$ ice restructures, opening some previously closed pores. The finally remaining CO, which is at most at a 1:5 mixing ratio with CO$_2$ in thin ices, stays trapped until the onset of CO$_2$ desorption between 65 and 75~K.

\subsection{H$_2$O vs CO$_2$ entrapment \label{disc-h2o}}

The most surprising outcome of this study is that prior to the onset of CO$_2$ desorption, pure CO$_2$ traps CO molecules more efficiently than comparable H$_2$O ices. This result may be due to differences in the CO$_2$ and H$_2$O ice binding environments, structures, or dynamics. We explore the possible contributions from each in this sub-section.

The CO binding energies to CO$_2$ and H$_2$O molecules in ice have been constrained through TPD experiments in which sub-monolayers of CO are deposited on compact, thermally annealed CO$_2$ and H$_2$O ices. The key result is that the desorption temperatures and therefore binding energies are the same for CO on CO$_2$ and CO on H$_2$O ice within experimental uncertainties \citep{Fayolle16,Cooke18}.  
By contrast, the CO desorption energy for CO deposited on a porous amorphous H$_2$O ice, constructed at 15~K through direct vapor deposition, is substantially higher than for CO deposited on an analogous CO$_2$ ice \citep{Cooke18}. This implies that water ice deposited at $\sim$10--15~K can host more CO in strongly bound sites (e.g. in micropores) than CO$_2$ ice deposited at the same temperature. This suggests that low-temperature H$_2$O ice is more porous than low-temperature CO$_2$ ice. Such a porous structure may be detrimental to CO retention in mixed ices if pore collapse is slow compared to CO desorption, but may be beneficial to capture CO into the ice in the first place.

\begin{figure}[h!]
  \centering
  \includegraphics[width=0.4\textwidth]{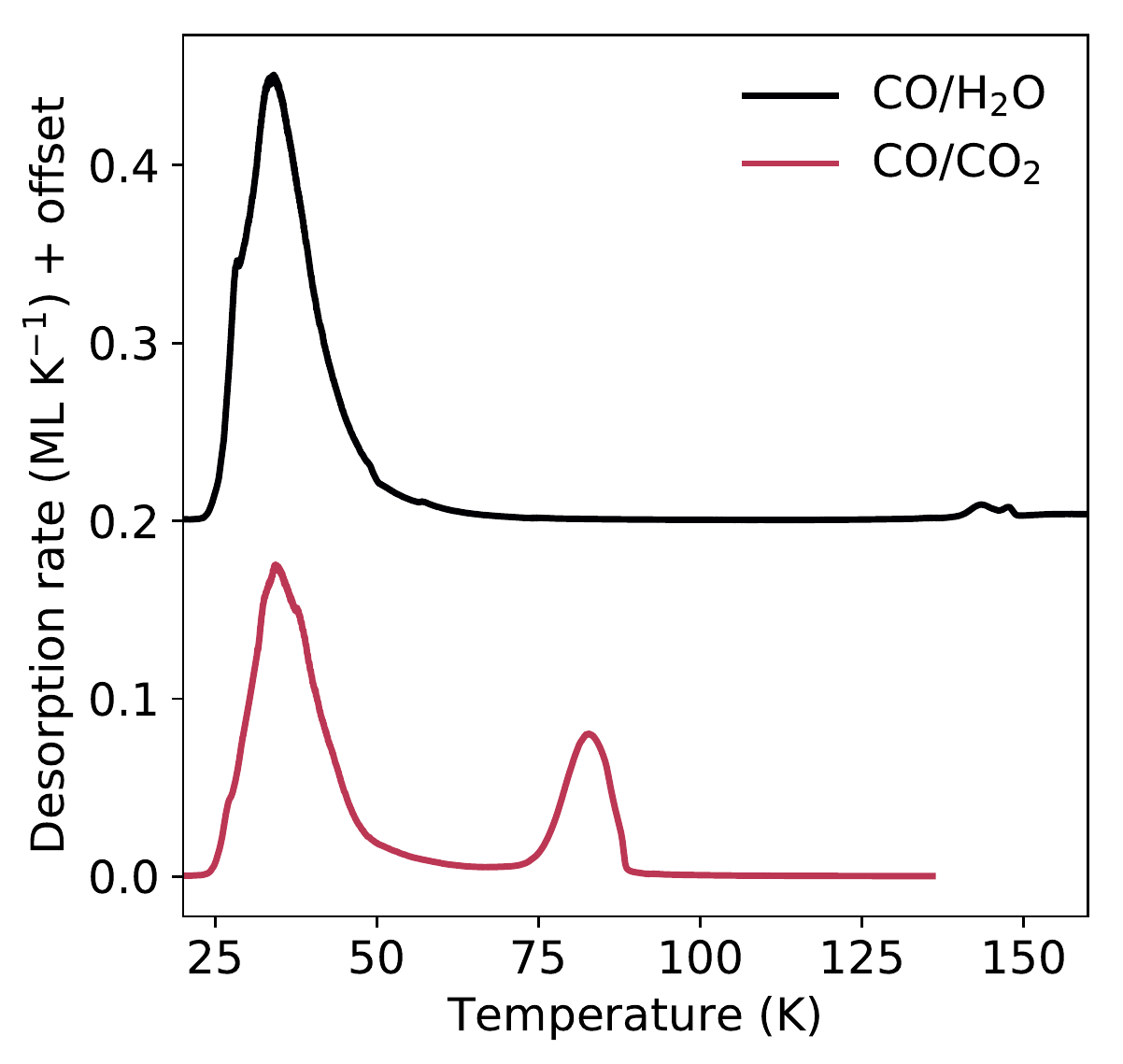}
 \caption{TPD spectra of CO desorbing off a CO$_2$ and H$_2$O ice respectively. In both cases the initial ice configuration was $\sim$3~ML of CO layered on top of $\sim$50~ML of CO$_2$ or H$_2$O, all deposited at 15~K. Note that in the case of CO$_2$, $\sim$15\% of the CO is retained, while in the case of H$_2$O only 5\% is.}
 \label{fig:co2-pore}
\end{figure}

To test whether H$_2$O ice is more efficient than CO$_2$ at entrapping CO molecules not already mixed into the ice, we performed two supplemental, layered ice experiments. In these experiments $\sim$3~ML of CO is deposited on top of $\sim$50~ML of CO$_2$ or H$_2$O ice. The substrate temperature was kept at 15~K during all depositions. Figure \ref{fig:co2-pore} shows that also in this scenario CO$_2$ traps substantially more of the deposited CO compared to the H$_2$O ice, similar to in the ice mixture experiments. Previous experiments on water ice have shown that the entrapment efficiency increases with ice porosity \citep{Ayotte01}. This comparison suggests that the difference between CO$_2$ and H$_2$O ice entrapment efficiencies is not mainly determined by different initial surface pore distributions.

Finally we consider the different ice restructuring processes in H$_2$O and CO$_2$ ices as a possible cause of the different entrapment efficiencies. Both water and CO$_2$ undergo phase changes between the onset of CO desorption around 30~K, and 70~K. In the case of water this is a phase change between low-density amorphous water and high-density amorphous water \citep[e.g.][]{Collings03b}, while in the case of CO$_2$ it is a drawn-out phase change from amorphous CO$_2$ to crystalline CO$_2$ ice \citep{Isokoski13,He2018}. The latter transformation may be more efficient in trapping CO in the ice. CO$_2$ ice restructuring may also begin earlier than the water phase change due to higher CO$_2$ ice mobility. Indeed \citet{Escribano13} and \citet{He2018} show that some CO$_2$ restructuring begins already at 20~K, and CO$_2$  may thus begin to trap CO before the onset of CO desorption.

\subsection{Astrophysical implications \label{disc-astro}}

During planet formation, the distribution of abundant volatiles between the solid and gas phases regulates the final compositions of planetary cores, hydrospheres and atmospheres.  Entrapment of hyper-volatiles such as CO and N$_2$, in less volatile ice matrices may change this distribution substantially, maintaining e.g. CO far interior to its expected condensation front or snowline. In this paper we have shown that  water ice is not the only common interstellar ice that can entrap molecules. CO$_2$, generally the second most abundant ice constituent \citep{Oberg11c,Boogert15}, may also trap CO during planet formation, and at a higher efficiency than H$_2$O in thin ices, as long as the ice is too cold for CO$_2$ sublimation. Since CO$_2$ is an abundant ice constituent, the identified difference in entrapment efficiency of $\sim$2 implies that H$_2$O and CO$_2$ ice may trap comparable quantities of CO in the ISM and in planet forming disks if H$_2$O and CO$_2$ reside in different ice phases. This is a big 'if' since in star forming regions most CO$_2$ is mixed in with water ice. As already introduced in \S\ref{sec:intro} there are several reasons why the case may be different in disks. First, the two comets studied in detail seem to present H$_2$O and CO$_2$ sublimations from different comet locations \citep{Mumma11,Hoang17,Gasc17}. While some of this difference can be explained by differences in sublimation temperatures between H$_2$O and CO$_2$, a substantial amount of CO$_2$ would be expected to co-desorb with H$_2$O if they were perfectly mixed together \citep{Fayolle11a}. Second there is theoretical evidence for substantial conversion of CO into CO$_2$ in protoplanetary disks based on astrochemical models \citep{Schwarz18} and low CO/H$_2$ ratios \citep{Bergin13}. Such a conversion on the surfaces of grains would naturally result in a CO$_2$:CO ice phase separate from that of H$_2$O.

Assuming then that all H$_2$O and CO$_2$ ice are available for CO entrapment, we can ask how much CO could be retained in ice grains in disks interior to the nominal CO snowline. Our study suggests that a substantial fraction could be present in CO$_2$ ice alone between the CO and CO$_2$ snowlines. In star-forming clouds, CO$_2$ ice abundances are as high as $2.6\times10^{-5}$ per $n_{\rm H}$ \citep{Boogert15}. Adopting our maximum observed entrapment CO$_2$:CO mixing ratio of 5:1, the maximum CO abundance that could be trapped in CO$_2$ ice is $5.2\times10^{-6}$ per $n_{\rm H}$. This is comparable to estimates of CO abundances in some protoplanetary disks \citep{Bergin13,Schwarz16,Cleeves16c,Miotello17,Long17}. It is important to note, however, that not all or even not most CO$_2$ may be present as pure ice, and future studies should explore CO entrapment in H$_2$O:CO$_2$ mixed ices, similarly to those commonly identified in the ISM.

Our experiments only probes entrapment of CO. Whether CO$_2$ also efficiently entraps N$_2$, CH$_4$, O$_2$ and noble gases remains to be explored. The proposed entrapment mechanism due to pore collapse around 30~K should be species agnostic, but it is possible that less volatile molecules than CO, e.g. N$_2$, begins to desorb prior to the onset of pore collapse. If this is the case, selective CO$_2$ entrapment may provide an alternative explanation for the low N$_2$/CO ratio observed in most comets, most recently in 67P \citep{Rubin15}, than preferential CO clathration in water ice \citep{Mousis12,Lectez15}.

\section{Conclusion}

Entrapment of hyper-volatiles in less volatile ice matrices changes the gas-phase and ice-phase molecular and elemental compositions during planet formation.  In this experimental study we show that CO$_2$ ice is a potential trapping agent in the ISM and in planet forming disks. CO$_2$ ice surprisingly traps CO more efficiently than analogous water ices. This may be explained by an earlier onset of pore collapse in CO$_2$ ice compared to H$_2$O ice, trapping a majority of CO molecules prior to the onset of CO desorption. CO$_2$ entrapment efficiencies are 40--60\% for the large majority of explored ice compositions and thicknesses, indicating that substantial quantities of CO may be trapped in CO$_2$ ice. This discovery increases the theoretical amount of CO that can be retained in ice in between the CO$_2$ and CO snowlines in disks. It may also help to explain the correlation of CO and CO$_2$ outgassing from 67P.

\acknowledgments 

AS is grateful to Jennifer Bergner for assistance with constructing data analysis scripts. KI\"O acknowledges funding from the Simons foundation through the Simons Collaboration on Origins of Life (SCOL) Investigator Award (\#321183)

\end{document}